\documentclass[%
 reprint,
 amsmath,amssymb,
 aps,
]{revtex4-2}

\usepackage{graphicx}% Include figure files
\usepackage{dcolumn}% Align table columns on decimal point
\usepackage{bm}% bold math
\begin{document}

\preprint{APS/123-QED}

\title{Symmetry-protected higher-order exceptional points in staggered flatband rhombic lattices}% Force line breaks with \\
\thanks{These authors contributed equally.\\
$\dagger$ xiashiqiang@htu.edu.cn\\
$\ddag$ phyzhxd@gmail.com\\
$\sharp$ tanya@nankai.edu.cn}%
%\author{}

\author{Yingying~Zhang$^{1*}$, Shiqiang~Xia$^{1*\dagger}$, Xingdong~Zhao$^{1\ddag}$, Lu~Qin$^{1}$, Xuejing~Feng$^{1}$, Wenrong~Qi$^{1}$, Yajing~Jiang$^{1}$,\\
Hai~Lu$^{1}$, Daohong~Song$^{2}$, Liqin~Tang$^{2\sharp}$, Zunlue~Zhu$^{1} $ and~Yufang~Liu$^{1}$}

\affiliation{$^1$ School of Physics, Henan Normal University, Xinxiang, Henan 453007, China\\
$^2$ The MOE Key Laboratory of Weak-Light Nonlinear Photonics, TEDA Applied Physics Institute and School of Physics, Nankai University, Tianjin 300457, China
}

%\author{Yingying Zhang,\authormark{1,\#}, Shiqiang Xia,\authormark{1,3,\#}, Xingdong Zhao,\authormark{1,4}, Lu Qin,\authormark{1}, Xuejing Feng,\authormark{1}, Wenrong Qi,\authormark{1}, Yajing Jiang,\authormark{1}, Hai Lu,\authormark{1},Zunlue Zhu,\authormark{1} and Yufang Liu,\authormark{1}}
% \altaffiliation[Also at ]{School of Physics, Henan Normal University, Xinxiang, Henan 453007, China\\}%Lines break automatically or can be forced with \\


%\author{Daohong Song,\authormark{2}, Liqin Tang,\authormark{2,5}}
%\affiliation{


\date{\today}% It is always \today, today,

\begin{abstract}
Higher-order exceptional points (EPs), which appear as multifold degeneracies in the spectra of non-Hermitian systems, are garnering extensive attention in various multidisciplinary fields. However, constructing higher-order EPs still remains as a challenge due to the strict requirement of the system symmetries. Here we demonstrate that higher-order EPs can be judiciously fabricated in $\mathcal{PT}$-symmetric staggered rhombic lattices by introducing not only on-site gain/loss but also non-Hermitian couplings. Zero-energy flatbands persist and symmetry-protected third-order EPs (EP3) arise in these systems owing to the non-Hermitian chiral/sublattice symmetry, but distinct phase transitions and propagation dynamics occur. Specifically, the EP3 arises at the Brillouin zone (BZ) boundary in the presence of on-site gain/loss. The single-site excitations display an exponential power increase in the $\mathcal{PT}$-broken phase. Meanwhile, a nearly flatband sustains when a small lattice perturbation is applied. For the lattices with non-Hermitian couplings, however, the EP3 appears at the BZ center. Quite remarkably, our analysis unveils a dynamical delocalization-localization transition for the excitation of the dispersive bands and a quartic power increase beyond the EP3. Our scheme provides a new platform towards the investigation of the higher-order EPs, and can be further extended to the study of topological phase transitions or nonlinear processes associated with higher-order EPs.

%\begin{description}
%\end{description}
\end{abstract}

\maketitle

%\tableofcontents

\section{\label{sec:level1}Introduction}

Exceptional points (EPs), singularities at which eigenvalues and eigenvectors simultaneously coalesce, have attracted considerable interest in the non-Hermitian physics~\cite{feng2017non,el2018non,ozdemir2019parity,miri2019exceptional,parto2021non,bergholtz2021exceptional,PhysRevLett.126.230402}. For instance, EPs in the parity–time ($\mathcal{PT}$)-symmetric optical arrangements~\cite{PhysRevLett.100.103904,PhysRevLett.101.080402} have provided an excellent platform for exploring intriguing phenomena such as the $\mathcal{PT}$-symmetric lasers~\cite{hodaei2014parity,feng2014single} and topological phase transitions~\cite{PhysRevLett.115.200402,PhysRevLett.120.146402,xia2021nonlinear}. In particular, recently, there has been a great deal of interest in proposing various methods for constructing higher-order (greater than second order) EPs. The eigenvalue shift near the higher-order EPs follows the $\epsilon ^{1/N} $ ($N$ is the order of the EPs) power law of the external perturbation $\epsilon$, therefore leading to higher sensitivity of resonant optical structures to small external disturbances ~\cite{chen2017exceptional,hodaei2017enhanced,hokmabadi2019non,duggan2022limitations}. Another interesting thing is that optical systems with $\mathcal{PT}$ symmetry can show additional symmetries, such as time-reversal symmetry and chiral symmetry (CS), making the higher-order EPs much more abundant~\cite{longhi2018pt,PhysRevLett.127.186601,PhysRevLett.127.186602}. In fact, these symmetries ramify and show new features in non-Hermitian systems due to the fact that the Hamiltonian $H\ne H^{\dagger } $. A crucial example is CS which satisfies $CHC^{-1} =-H$. The CS coincides with sublattice symmetry (SLS) ($SH^{\dagger } S^{-1} =-H$) in the presence of Hermiticity. Obviously, they are not equivalent for non-Hermitian Hamiltonians~\cite{kawabata2019topological,PhysRevX.9.041015,PhysRevB.103.235130,PhysRevB.103.014111}.

The study of EPs has also been extended to flatband systems. A flatband is dispersionless throughout the Brillouin zone (BZ) and it provides a unique setting for exploring anomalous magnetic phases and strongly correlated states of matter~\cite{PhysRevLett.106.236803,PhysRevB.83.220503,balents2020superconductivity,wang2020localization}. In flatband systems, diffraction is suppressed due to destructive interference of the Bloch wave functions, resulting in typically compact localized states (CLSs) in real space~\cite{PhysRevLett.114.245503,PhysRevLett.114.245504,PhysRevLett.120.097401,leykam2018perspective,tang2020photonic,poblete2021photonic}. The most direct way to form $\mathcal{PT}$-symmetric flatband system is to introduce on-site gain/loss to the lattices. Recently, the $\mathcal{PT}$-symmetric flatband photonic waveguide arrays with tailored on-site gain/loss distributions have been experimentally created using laser-written technique~\cite{weimann2017topologically,PhysRevLett.123.183601}. On the other hand, the non-Hermitian couplings describe a situation where the mode amplitude undergoes gain/loss while hopping between sites~\cite{PhysRevA.99.033810,PhysRevLett.120.093901,longhi2018invited}. It has shown that the $\mathcal{PT}$-symmetric frustrated lattices with fine-tuned non-Hermitian couplings can also give rise to flatbands as well as EPs~\cite{PhysRevB.96.064305}. More importantly, recent progress has demonstrated that even the higher-order EPs can exist in the $\mathcal{PT}$-symmetric flatband lattices~\cite{PhysRevLett.127.186601,PhysRevLett.127.186602,xia2021higher}. Nevertheless, due to the band touching (vanishing bandgap) between flat and dispersive bands, so far, realizations of the non-zero valued higher-order EPs have been limited to flatband systems with non-Hermitian couplings~\cite{PhysRevLett.121.263902,rhim2020quantum,ge2018non,PhysRevB.102.245144}. Moreover, the distinct features of the two aspects of non-Hermiticity (on-site gain/loss and non-Hermitian couplings) as well as the dynamical behaviors near the phase transitions, also remain elusive.

In this work, we demonstrate that a higher-order EP (EP3) can occur in staggered rhombic lattices by introducing the diagonal $\mathcal{PT}$ symmetry (with on-site gain/loss) and the off-diagonal $\mathcal{PT}$ symmetry (with non-Hermitian couplings) to the system. These $\mathcal{PT}$-symmetric lattices possess both the non-Hermitian CS and SLS, thus leading to the presence of zero-energy flatbands and non-zero EP3. However, the EP3 and the propagation dynamics of the lattices, unexpectedly, show strikingly different properties. For the diagonal $\mathcal{PT}$-symmetric lattices, a EP3 arises at the BZ boundary. We find that the single-site excitations of different sublattices display identical exponential power increases in the $\mathcal{PT}$-broken phase. By contrast, a EP3 is generated in the BZ center for the off-diagonal $\mathcal{PT}$-symmetric lattices. Intriguing propagation dynamics, such as a dynamical delocalization-localization transition for the excitation of dispersive bands and a quartic power increase beyond the EP3, are observed. For both diagonal and off-diagonal $\mathcal{PT}$-symmetric systems, the eigenvalue separations exhibit an $\sim \epsilon ^{1/3}$ dependence in the vicinity of the EP3. In particular, we show that a nearly flatband sustains for the diagonal $\mathcal{PT}$-symmetric lattices in a small range of perturbations. 

\section{The diagonal $\mathcal{PT}$-symmetric flatband rhombic lattices}

\begin{figure}
	\centering
	\includegraphics[width=0.8\columnwidth]{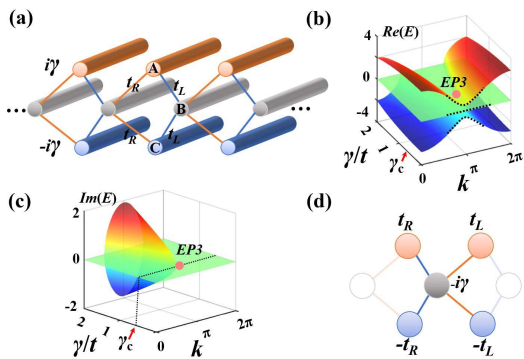}
	\caption{(a) Schematic of the diagonal $\mathcal{PT}$ -symmetric photonic rhombic lattices formed by waveguide arrays without non-Hermitian coupling. The unit cell consists of three sites: two edge sites with gain $\gamma$ ($\bf{A}$, red) or loss -$\gamma$ ($\bf{C}$, blue) and one central site with a neutral imaginary part ($\bf{B}$, gray). The staggered coupling coefficients are $t_L=t(1+g)$ and $t_R=t(1-g)$, where $0\le g< 1$ and $t$ is the coupling coefficient of uniform lattices ($g=0$). For simplicity, we set the lattice period $l=1$. (b, c) Calculated real (b) and imaginary (c) parts of the spectrum as a function of $\gamma/t$ for $g=0.2$. The red circles represent the position of the EP3 and the corresponding critical value of phase transition is $\gamma_c/t$=0.56. (d) Field distribution of a flatband CLS, which occupies two unit cells. The characters show the corresponding amplitudes.
		\label{figure1}}
\end{figure}

\begin{figure*}[htbp]
	\centering
	\includegraphics[width=1.7\columnwidth]{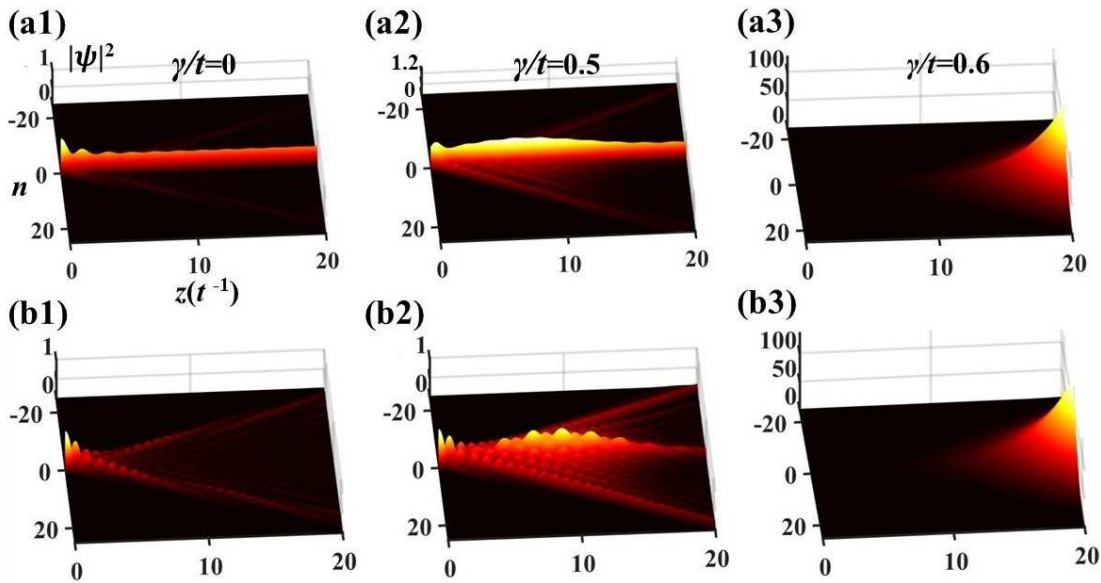}
	\caption{Intensity profile $\left | \psi  \right | ^{2} $ of a single-site $\bf{A}$ (a1-a3) or $\bf{B}$ (b1-b3) excitation. (a1) The $\bf{A}$ sublattice excitation for Hermitian case ($\gamma/t=0$) exhibits a suppression of diffraction due to the excitation of flatband CLSs. (a2) The input remains conserved below the EP3 threshold ($\gamma/t=0.5$). (a3) The input leads to an exponentially growing total intensity in the $\mathcal{PT}$-broken phase ($\gamma/t=0.6$). (b1) The $\bf{B}$ site excitation for Hermitian case ($\gamma/t=0$), showing a typically discrete diffraction pattern. (b2) The input excites both the flatband and dispersive bands, producing discrete diffraction as well as a residual localized component ($\gamma/t=0.5$). (b3) The input shows same exponential growth of total intensity as in (a3) ($\gamma/t=0.6$). The propagation length is $z$=20 $t^{-1}$. Note the different scales of the vertical axis in these panels, though the initial amplitudes of the excitation all equal 1.
		\label{figure2}
	}
\end{figure*}

\textbf {\textit{Symmetry-protected flatband and EP3}}--We start from introducing the diagonal $\mathcal{PT}$-symmetric staggered photonic rhombic lattices with on-site gain/loss to elucidate the lattice symmetry, the formation of flatband and the EP3. Figure~\ref{figure1}(a) shows the lattice structure with each unit cell consisting of an edge site $\bf{A}$ with gain ${\gamma}$, an edge site $\bf{C}$ with loss ${-\gamma}$, and a central neutral site $\bf{B}$. The sublattice $\bf{B}$ are fourfold connected with the nearest neighbors, while the sublattices $\bf{A}$ and $\bf{C}$ have twofold connections with surrounding sites. Experimentally, such non-Hermitian lattices can be obtained using the optical induced method or the femtosecond laser-writing technique ~\cite{weimann2017topologically,PhysRevLett.123.183601,xia2021nonlinear}. The couplings depend on the spacing of the neighbor waveguides, thus a staggered arrangement of the distance between the waveguides generates staggered couplings. Meanwhile, the on-site gain/loss can be generated by inscribing the waveguides with tailored loss distributions. The staggered couplings are $t_{L} =t(1+g)$ and $t_{R} =t(1-g)$, where $t_{L} (t_{R})$ is the coupling on the left (right) side of the site $\bf{B}$. Note that $t_{L} (t_{R})$ is real and represents conserved (Hermitian) coupling. In the tight binding model, the Hamiltonian can be written 
\begin{eqnarray}\label{Hamiltonian 1}
{H}=&& {\sum_{n}} \big ( t_Lb_n^{\dagger}a_{n}+t_Rb_n^{\dagger}a_{n+1}+t_Lb_n^{\dagger}c_{n}+t_Rb_n^{\dagger}c_{n+1}\notag\\
&&+{\rm H.c.}\big)+i\gamma\left(a_n^{\dagger}a_{n}-c_n^{\dagger}c_{n}\right),
\end{eqnarray}
where $a_{n}^{\dagger}$ ($a_{n}$), $b_{n}^{\dagger}$ ($b_{n}$) and $c_{n}^{\dagger}$ ($c_{n}$) are the creation (annihilation) operators in the $n$-th unit cell on the $\bf{A}$, $\bf{B}$ and $\bf{C}$ sites, respectively. Transforming $H$ into momentum space as $H=\sum_{k}\psi_{k}^{\dagger} H_{k} \psi _{k} $ with $\psi _{k} =\left ( a_{k},b_{k},c_{k} \right ) ^{T} $ , one can get the momentum space Hamiltonian
\begin{eqnarray}\label{Hamiltonian 11}
{H_{k}}=\begin{pmatrix}
i\gamma & t_L+t_R{e^{ik} }   & 0\\
t_L+t_R{e^{-ik} }& 0 & t_L+t_R{e^{-ik} }\\
0& t_L+t_R{e^{ik} } & -i\gamma
\end{pmatrix}.
\end{eqnarray}
The diagonal terms of $H _{k}$ describe the propagation constant and on-site gain/loss, whereas the the off-diagonal terms describe the lattice couplings. Such a Hamiltonian possesses $\mathcal{PT}$ symmetry $(PT){H_{k}}( PT)^{-1}={H_{-k}}$, and more importantly, a CS as well as a SLS satisfying $CH_kC^{-1} =-H_k$ and $SH_k^{\dagger } S^{-1} =-H_k$, where the unitary operator $C$ and $S$ are defined as 
\begin{eqnarray}
{C}=\begin{pmatrix}
0&  0& 1\\
0&  -1& 0\\
1&  0&0
\end{pmatrix},\
{S}=\begin{pmatrix}
1&  0& 0\\
0&  -1& 0\\
0&  0&1
\end{pmatrix}.
\end{eqnarray}
By calculating the eigenvalues of $H_k$, one can easily get the three corresponding eigenvalues:
\begin{gather}
{{E_{0}(k)=0}},\nonumber\\
{E_{\pm 1} \left ( k \right ) =\pm \sqrt{-\gamma ^{2} +4t^{2} \left ( 1+ {\rm cos}k \right ) +4t^{2}g^{2} \left ( 1-{\rm cos}k \right )  }}.
\end{gather}

In the Hermitian case ($\gamma /t=0$), the CS/SLS symmetry requires all the energy eigenvalues appear in pairs, namely, the three-band system must have a zero-energy ﬂatband $E_0=0$ and two dispersive bands with opposite energies $E_1=-E_{-1}$~\cite{PhysRevB.96.161104,PhysRevB.97.045120}. Figures~\ref{figure1}(b) and ~\ref{figure1}(c) show the spectrum as a function of $\gamma /t$ for $g$=0.2. It can be clearly seen that the Hermitian spectrum in the limit $\gamma /t=0$ has three bands: a zero-energy flatband located between two symmetric dispersive bands (dotted lines in Fig.~\ref{figure1}(b)). Unlike in uniform lattices where all the bands touch each other at $k=\pi $ ~\cite{xia2020observation,kremer2020square,PhysRevResearch.2.033127}, here the band degeneracy is lifted and two symmetric gaps with gap width $E_{gap} =2\sqrt{-\gamma ^{2}+8t^{2} g^{2}} $ emerge as a result of the staggered couplings. When the on-site gain/loss is introduced ($\gamma /t\ne 0$), the energy eigenvalues become complex. As mentioned above, the CS bifurcates into CS and SLS in the non-Hermitian case. However, we find both the CS and SLS preserve. Meanwhile, they retain $E_1=-E_{-1}$ and non-Hermitian zero mode occurs for the central band. As a consquence, the zero-energy flatband survives and the spectrum stays symmetric about the imaginary and real energy
axis of the complex energy plane. The band gaps decrease by increasing $\gamma/t$. More importantly, all the three eigenvalues coalesce at $E_{0}=0$ for the critical value $\gamma _{c} /t=2\sqrt{2} g=0.56$ (Figs.~\ref{figure1}(b) and ~\ref{figure1}(c)). Note that $\gamma_c$ depends on the value of $g$, i.e.,  the relative strength of the staggered couplings. The three eigenvectors also coalesce at this specific degenerate point, and have the form $\left [ \psi_{n}^{A} , \psi_{n}^{B} ,\psi_{n}^{C} ,\psi_{n+1}^{A} ,\psi_{n+1}^{C}  \right ]=\left [ 0.8,-0.2i,-0.8,1.2,-1.2 \right ]$. Accordingly, the spectrum is completely real for $\gamma < \gamma _{c} $, and a CS/SLS-protected EP3 emerges at $\gamma = \gamma _{c}  $. For $\gamma > \gamma _{c}  $, the system turns into the $\mathcal{PT}$-broken phase. Figure~\ref{figure1}(d) illustrates the irreducible CLS of the flatband. As  demonstrated in previous studies, CLSs are localized in a single unit cell in Hermitian rhombic lattices~\cite{Xia2021band,PhysRevLett.121.075502,mukherjee2015observation}. Here, in contrast, the CLS is confined in the five-site-cross configurations in two adjacent unit cells. Non-zero amplitudes appear in $\bf{A}$ and $\bf{C}$ sites and the on-site gain/loss remodulates the amplitude of the $\bf{B}$ site. The equal amplitude and opposite phase ensure the destructive interference between the two legs formed by the $\bf{A}$ and $\bf{C}$ sites.

\begin{figure}[t]
	%\centering
	\includegraphics[width=0.8\columnwidth]{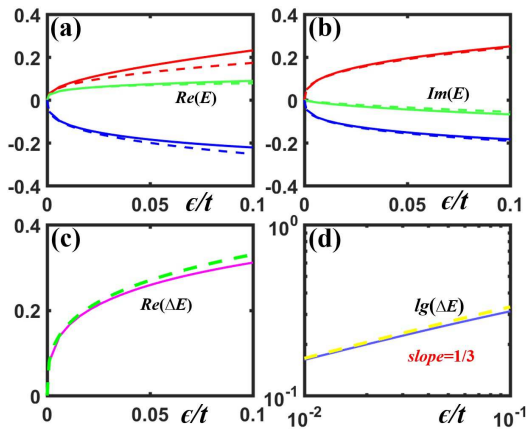}
	\caption{(a, b) Analytical (dashed lines) and numerical (solid lines) results for the real (a) and imaginary (b) parts of the  spectrum as a function of the detuning $\epsilon/t$ in the vicinity of EP3. Red lines, $E_{-1}$; green lines, $E_{0} $; blue lines, $E_{1} $. (c) Corresponding results for $E_{0} -E_{1} =\bigtriangleup E$, exhibiting a cube-root behavior. (d) The results from (c) on a logarithmic scale. The slope of about 1/3 confirms the cube-root response. 
		\label{figure3}}
\end{figure}

\textbf {\textit{Propagation dynamics for single-site excitation}}--To verify our analysis, we simulate the light propagation dynamics for initial excitations localized to either an $\bf{A}$ or $\bf{B}$ site. A sufficiently large number of lattice sites (101 unit cells) is considered. The input is $a_n(0) $ or $b_n\left ( 0 \right ) =\left ( \sqrt{\pi }/\sigma   \right ) e^{-\left ( n^{2} /\sigma ^{2}  \right ) }\left ( \sigma =0.5 \right )$, i.e., only the central single site is excited. Figure~\ref{figure2} shows the evolution of the beam intensity $P\left ( z \right ) =\sum_{n}\left | \psi _{n}  \left ( z \right ) \right | ^{2} =\sum_{n}\left ( \left | a _{n}  \left ( z \right ) \right | ^{2}+\left | b _{n}  \left ( z \right ) \right | ^{2}+\left | c _{n}  \left ( z \right ) \right | ^{2} \right )$, where $z$ is an effective propagation distance. For comparison, we first consider the result for the Hermitian case at $\gamma /t=0$. A suppression of diffraction is observed for the excitation of $\bf{A}$ site (Fig.~\ref{figure2}(a)). The input excites a superposition of dispersive bands and flatband states. The energy is mainly distributed in the flatband, therefore most of the power remains confined to the initially excited site. When the system is operated below the $\mathcal{PT}$ threshold value ($\gamma /t=0.5$), part of the energy evolves into the dispersive bands though most of the energy still stays localized (Fig.~\ref{figure2}(a2)). It should be noted that, what is conserved here is the quasi-energy $Q\left ( z \right ) =\sum_{n}\psi \left ( -n,z \right )  \psi ^{\ast } \left ( n,z \right )$ as opposed to the actual power $P(z)$. The presence of the gain/loss results in exponential amplification $\left [ P\left ( z \right )\sim e^{z}   \right ]  $ of the total power when the non-Hermitian parameter is tuned to $\gamma/t$=0.6 (Fig.~\ref{figure2}(a3)) (see Appendix~\ref{APA}). A different scenario is found when a localized $\bf{B}$ site is excited. In Fig.~\ref{figure2}(b1), the input generates a conventional discrete diffraction pattern with two ballistically expanding lobes for the Hermitian lattices. In this case, only the dispersive bands are excited since no flatband energy is distributed in the $\bf{B}$ sublattice (cf. Fig.~\ref{figure1}(d)). Nevertheless, the dispersive bands and flatband states, including the EP3 states are excited in the $\mathcal{PT}$ symmetry phase $\gamma/t$=0.5 (Fig.~\ref{figure2}(b2)). The resulting diffraction consists of a discrete pattern and conserved total power as required for a $\mathcal{PT}$-symmetric lattice in the unbroken phase. Furthermore, there is a residual localized component due to the excitation of the flatband. With the further increase of $\gamma/t$, the energy is also obviously amplified when the system is operated in the $\mathcal{PT}$ symmetry broken phase for $\gamma/t$=0.6 (Fig.~\ref{figure2}(b3)). Interestingly, we find that the power growth is the same as that in Fig.~\ref{figure2}(a3).

\textbf {\textit{Bifurcations of eigenvalues around the EP3}}--In practical implementations, the propagation of photons is subject to inevitable perturbations. As mentioned above, the eigenvalues bifurcation characteristics of the higher-order EPs can enhance the response of the optical structure to small external disturbances. In the following, we demonstrate the eigenvalues evolution of our ternary $\mathcal{PT}$-symmetric photonic structure around the EP3 and show how a small lattice perturbation affects the wave propagation dynamics.

Note that the perturbation could be introduced anywhere along the diagonal of the matrix $H_k$. For simplicity and without loss of generality, we suppose that the external perturbation $\epsilon $ is imposed on the sublattice $\bf{A}$. 
%%%%%%%%%%%%%%%%%%%%%%%%Fig4%%%%%%%%%
\begin{figure}[b]
	\centering
	\includegraphics[width=0.90\columnwidth]{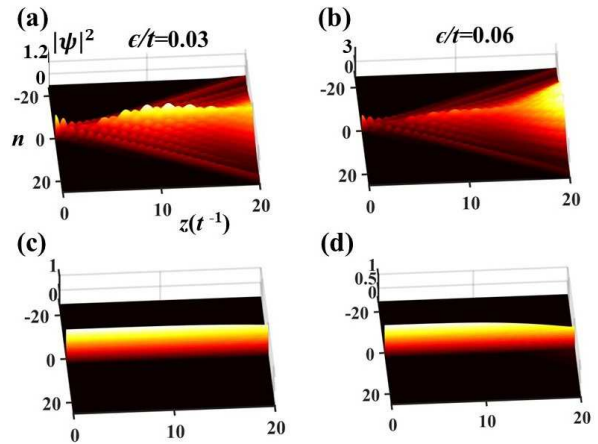}
	\caption{(a, b) Intensity profile for the $\bf{B}$ site excitation in the presence of small perturbations $\epsilon/t=0.03$ and  $\epsilon/t=0.06$, respectively. For $\epsilon/t=0.03$, a similar pattern compared with Fig.~\ref{figure2}(b2) is obtained. (c, d) Same as (a, b) but for the excitation of a flatband CLS. The CLS can stay well localized despite the presence of the lattice perturbations.
		\label{figure4}}
\end{figure}
The energy bands can be analytically obtained through solving the cubic characteristic equation (see Appendix~\ref{APB}). Figures~\ref{figure3}(a) and ~\ref{figure3}(b) illustrate analytical and numerical solutions for the real and imaginary parts of the eigenvalues as a function of $\epsilon /t$. The small perturbation ranges from $10^{-10} $ to $10^{-1} $. The analytical solutions show a good agreement with numerical results. The perturbation drives the system out of the degenerate state and three complex eigenvalues are obtained. To have a quantitative comparison, we calculate the numerical eigenvalue difference of the upper band in Fig.~\ref{figure3}(a) for $\epsilon /t=0.03$ and $\epsilon /t=0.06$, which are 0.1 and 0.15, respectively. The difference between the lower two eigenvalues ($E_{0}$ and $E_{1} $) is plotted in Fig.~\ref{figure3}(c). Obviously, the eigenvalues splitting increases along with the strength of the perturbation. By considering the logarithmic behavior of this curve, we find that the slope is about 1/3, which is exactly the characteristic of an EP3 (Fig.~\ref{figure3}(d)). Another intriguing feature is that both the real and imaginary parts of the central band change slightly, especially for $\epsilon /t< 0.03$. This indicates that a nearly flatband can be sustained in such a range of perturbations.

The perturbation on the spectrum can also be reflected by light propagation dynamics. For $\epsilon /t=0.03$, one can find a similar pattern for the excitation of $\bf{B}$ sublattice in Fig.~\ref{figure4}(a), which is compared with that in Fig.~\ref{figure2}(a2). An increase of the lattice perturbation ($\epsilon /t=0.06$) results in an enhancement of wave intensity in Fig.~\ref{figure4}(b). The input behaves differently for the excitation of the flatband CLSs. An interesting property of flatband is that, despite the non-zero couplings between lattice sites, propagation of flatband eigenstates is completely suppressed. Therefore, if we excite five sites according to Fig.~\ref{figure1}(d), we can observe a completely localized state with $z$-invariant intensity. In Fig.~\ref{figure4}(c), a CLS excitation can stay well localized in the presence of $\epsilon /t=0.03$, confirming that the flatband can preserve in this case. Moreover, the input keeps nearly localized even when the lattice detuning is increased up to $\epsilon /t=0.06$ as shown in Fig.~\ref{figure4}(d).

\section{The off-diagonal $\mathcal{PT}$-symmetric flatband rhombic lattices}
\begin{figure}
	\centering
	\includegraphics[width=0.8\columnwidth]{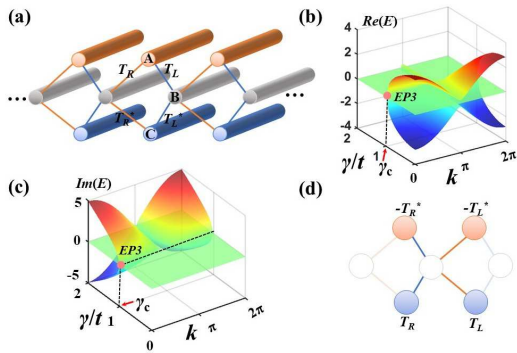}
	\caption{ (a) Schematic of the off-diagonal $\mathcal{PT}$-symmetric photonic rhombic lattices formed by waveguide arrays with non-Hermitian coupling. The complex-coupling coefficients are $T_L=t_L+i\gamma$, $T_L^{*}=t_L-i\gamma$, $T_R=t_R+i\gamma$, and $T_R^{*}=t_R-i\gamma$, respectively. Note that the real parts ($t_L$ and $t_R$) are the same as in Fig.~\ref{figure1}(a), which represent conserved staggered couplings. (b, c) Calculated real (b) and imaginary (c) parts of the spectrum as a function of $\gamma/t$ for $g=0.2$. The EP3 forms at $k=0$ and the critical value of the phase transition is $\gamma_c=t$. (d) Field distribution of the flatband CLS, which also occupies two unit cells but only distributes in $\bf{A}$ and $\bf{C}$ sites.
		\label{figure5}}
\end{figure}

\textbf {\textit{Symmetry-protected flatband and EP3}}--The most intriguing peculiarity of our scheme is that the non-Hermiticity can also be introduced by employing the non-Hermitian couplings. Such a system also possesses an EP3 and a flatband but exhibits different features. As illustrated in Fig.~\ref{figure5}(a), we maintain the geometric symmetry of the lattices in Fig.~\ref{figure1}(a) and perform imaginary processing on the coupling parameters. $T_L=t_L+i\gamma$, $T_L^{*}=t_L-i\gamma$, $T_R=t_R+i\gamma$, and $T_R^{*}=t_R-i\gamma$ are the complex-valued coupling coefficients on the left (right) side of the site $\bf{B}$, respectively. These complex couplings can be regarded as the phase of a hopping amplitude, which can be experimentally realized by embedding amplifying/lossy media between adjacent waveguides or periodically modulate the on-site gain/loss amplitude ratio~\cite{PhysRevB.96.064305,PhysRevA.89.013848,PhysRevA.93.022102,PhysRevB.92.094204,ding2021non}. Light propagates in engineered lattices with segmented regions of alternating gain/loss and index contrast regions. The effective gain/loss is sensitive to the relative phase between neighboring waveguides, leading to stronger or weaker amplifications. In this case, the Hamiltonian can be written as
%\begin{small}
\begin{eqnarray}\label{Hamiltonian 111}
{H}=&&{\textstyle \sum_{n}} \big ( T_Lb_n^{\dagger}a_{n}+T_Rb_n^{\dagger}a_{n+1}+T^*_Lb_n^{\dagger}c_{n}+T^*_Rb_n^{\dagger}c_{n+1}\notag\\
&&+{\rm H.c.}\big ).
\end{eqnarray}
%\end{small}
The corresponding Hamiltonian in momentum space is 
\begin{eqnarray}\label{Hamiltonian2}
%\begin{small}
{H_{k}}=\begin{pmatrix}
0 & T_L+T_R{e^{ik} }   & 0\\
T_L+T_R{e^{-ik} }& 0 & T^*_L+T^*_R{e^{-ik} }\\
0& T^*_L+T^*_R{e^{ik} } &0
\end{pmatrix},
%\end{small}
\end{eqnarray}
which satisfies the $\mathcal{PT}$ symmetry $\left ( PT \right ) H_{k} \left ( PT  \right )^{-1} =H_{-k}$. Interestingly, one can find that such a arrangement also possesses CS and SLS. Here, the unitary operator $C$ and $S$ can be set as 
\begin{eqnarray}
{C}=\begin{pmatrix}
1&  0& 0\\
0&  -1& 0\\
0&  0&1
\end{pmatrix},\
{S}=\begin{pmatrix}
0&  0& 1\\
0&  -1& 0\\
1&  0&0
\end{pmatrix}.
\end{eqnarray}
The energy eigenvalues of Hamiltonian (\ref{Hamiltonian2}) are
\begin{gather}
	{E_{0}(k)=0},\nonumber\\
	{E_{\pm 1} \left ( k \right ) =\pm 2\sqrt{\left ( t ^{2} -\gamma^{2}  \right )\left ( 1+\cos k \right ) +g^{2}t^{2}\left ( 1-\cos k \right )}}.
\end{gather}

\begin{figure*}[ht]
	\centering
	\includegraphics[width=1.7\columnwidth]{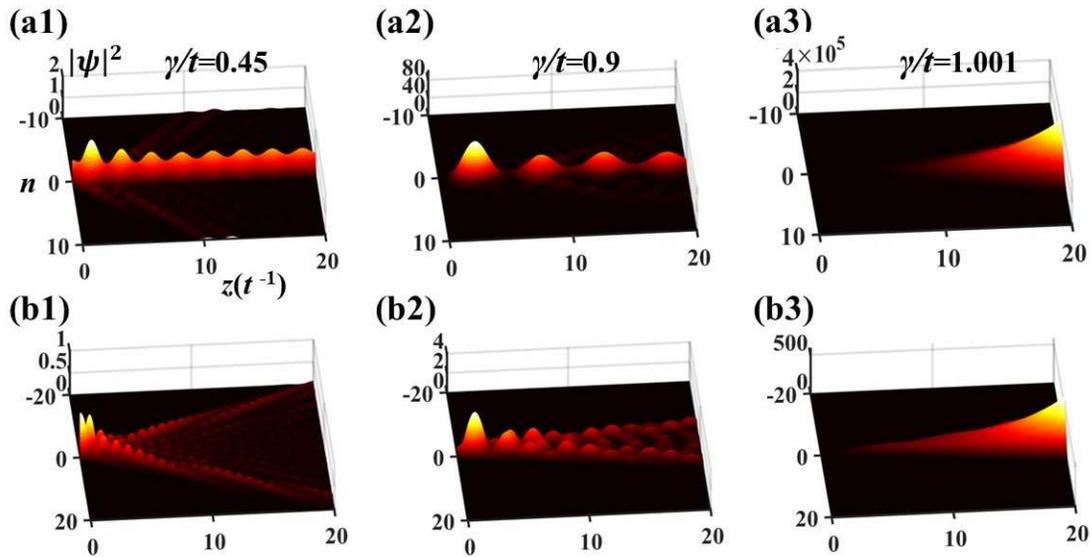}
	\caption{Same as Fig.~\ref{figure2}, but for the off-diagonal  $\mathcal{PT}$-symmetric lattices. (a1) The $\bf{A}$ sublattice excitation oscillates periodically below the EP3 threshold ($\gamma/t=0.45$). Most of the energy stay localized due to the excitation of the flatband. (a2) The quasi-energy is conserved and the oscillation period is doubled for $\gamma/t=0.9$. (a3) The input leads to quartic total intensity grow once the system is operated in the $\mathcal{PT}$ symmetry broken phase ($\gamma/t=1.001$). (b1) The $\bf{B}$ input only excites eigenstates of the dispersive bands and therefore experiences conventional diffraction. (b2) The spreading velocity diminishes with the increase of $\gamma/t$. (b3) The input becomes localized and the total intensity also show a quartic increase in the $\mathcal{PT}$ symmetry broken phase.
		\label{figure6}}
\end{figure*}

For direct comparison of the two non-Hermitian arrangements, we also set $g$=0.2 and plot the real and imaginary parts of the spectrum (as shown in Figs.~\ref{figure5}(b) and ~\ref{figure5}(c)). Due to the presence of CS and SLS, the central flatband still survives and the three bands distribute symmetrically. With the increase in $\gamma/t$, the bandgap tends to narrow and merge. In contrast, here, the $\mathcal{PT}$ symmetry broken transition occurs at the center of the BZ ($k=0$) and the critical value is $\gamma_c=t$, i.e., is invariable and independent of the value of $g$. The eigenmodes of the bands also coalesce into a degenerated one, namely, a CS/SLS-protected EP3 is generated. Above the EP3, the spectrum becomes complex. In addition, the irreducible flatband CLS also occupies two unit cells but the energy only distribute in $\bf{A}$ and $\bf{C}$ sites as shown in Fig.~\ref{figure5}(d).

\begin{figure}[h]
	\centering
	\includegraphics[width=0.8\columnwidth]{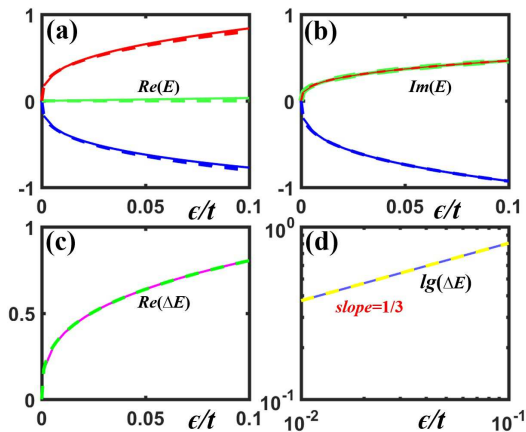}
	\caption{Same as Fig.~\ref{figure3}, but for the off-diagonal $\mathcal{PT}$-symmetric lattices.
		\label{figure7}}
\end{figure}

\textbf {\textit{Propagation dynamics for single-site excitation}}--In Fig.~\ref{figure6}, we illustrate the propagation dynamics of the system for the same initial excitations as in Fig.~\ref{figure2}. Similar patterns as depicted in Figs.~\ref{figure2}(a1) and ~\ref{figure2}(b1) can be obtained in the Hermitian limit $\gamma/t=0$. However, diverse propagation dynamics can be observed for $\gamma/t\ne0$. The input experiences periodically oscillations for the $\bf{A}$ site excitation due to the unfolding of the nonorthogonal Floquet-Bloch modes in $\mathcal{PT}$ symmetry phase $\gamma /t< 1$ (Figs.~\ref{figure6}(a1) and ~\ref{figure6}(a2)). The beat length, which is equal to ${L=2\pi /(\delta E)} $ (${\delta E} $ is the difference between the propagation constants of two modes), increases as $\gamma/t$ approaches the EP3~\cite{PhysRevLett.101.080402,makris2010pt,PhysRevA.95.053868}. As can be found in Fig.~\ref{figure6}(a2), the oscillation is enhanced and has a double period for  $\gamma/t=0.9$. The beat length goes to infinity as the EP3 is approached. Furthermore, we observe a sharp transition in Fig.~\ref{figure6}(a3). The input becomes localized and leads to a quartic power increase $\left [ P(z) \sim z^{4} \right ]$ in the $\mathcal{PT}$ symmetry broken phase (see Appendix~\ref{APA}).
The dynamics exhibited in Figs.~\ref{figure6}(b1-b3) is distinct from that in Figs.~\ref{figure6}(a1-a3). Transport is always ballistic in the $\mathcal{PT}$ symmetry phase ($\gamma /t< 1$) for the $\bf{B}$ site excitation. In this case, the input will never excite the flatband CLSs (cf. Fig.~\ref{figure5}(d)) and thus no localized component appears. More saliently, the excitation displays a transition from ballistic wave-packet spreading to dynamical localization, as the non-Hermitian parameter $\gamma /t$ is increased to the value at the EP3. In Figs.~\ref{figure6}(b1) and ~\ref{figure6}(b2), the excitation propagates bidirectionally along the lattice with the spreading speed $\nu \sim \sigma \left ( z \right ) /z$, where the position operator $\sigma^2 = {\textstyle \sum_{n}} n^{2} \psi_{n} ^{2} / {\textstyle \sum_{n}} \psi_{n} ^2$. $\nu $ decreases as $\gamma /t$ is increased and the wave-packet spreading decreases faster for large $\gamma /t$ (see Appendix~\ref{APC}). When $\gamma/t$ approaches the EP3, the diffraction is prevented. In fact, the largest velocity at which an excitation propagates along the lattice can also be described by the group velocity $\nu _{g} =Re\left \{ \left ( dE/dk \right ) _{k=\pi /2}  \right \} $. As the $\gamma /t$ is increased, the energy spectrum in Fig.~\ref{figure5}(b) undergoes a deformation, which changes the dispersion relation and results in a decrease of $\nu _{g}$~\cite{longhi2015robust,PhysRevB.103.054203}. This fact can be used for a direct observation of the EP: the propagation constants approach when the non-Hermitian parameter is increased to the value at the EP3. Though the total intensity in Fig.~\ref{figure6}(b3) is much smaller than that in Fig.~\ref{figure6}(a3), we find that it also follows a quartic power increase in the $\mathcal{PT}$ symmetry broken phase.

\textbf {\textit{Bifurcations of eigenvalues around the EP3}}--Similarly, we calculate the spectrum separation of the system in the presence of the external perturbations. The eigenvalues difference depicted in Fig.~\ref{figure7} also has a 1/3 power law of the external perturbation (Figs.~\ref{figure7}(c) and 
~\ref{figure7}(d)). Note that the changes of the actual value of both the dispersive bands and flatband are greater compared with that in Fig.~\ref{figure3}. For instance, the numerical eigenvalue difference in Fig.~\ref{figure7}(a) for $\epsilon /t=0.03$ and $\epsilon /t=0.06$ are 0.54 and 0.69, respectively. Besides, we find that the imaginary part of the central band is identical to the upper band, indicating that the flatband transforms into a dispersive band (Fig.~\ref{figure7}(b)).

\begin{figure}[h]
	\centering
	\includegraphics[width=0.90\columnwidth]{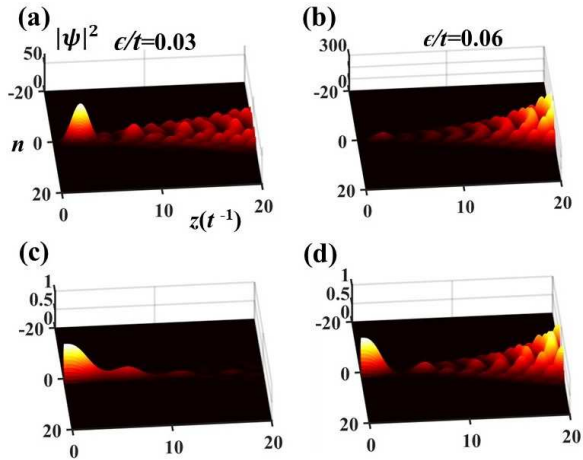}
	\caption{Same as Fig.~\ref{figure4}, but for the off-diagonal $\mathcal{PT}$-symmetric lattices.
		\label{figure8}}
\end{figure}

Figure~\ref{figure8} displays the corresponding propagation dynamics in the presence of the lattice perturbations. For $\bf{B}$ site excitation, the energy becomes obviously amplified for $\epsilon /t=0.06$  due to the complex spectrum (Fig.~\ref{figure8}(b)). More importantly, one can clearly find that the CLSs no longer exist even only a small perturbation is applied (Fig.~\ref{figure8}(c)) and the pattern in Fig.~\ref{figure8}(d) is similar to Fig.~\ref{figure8} (b). These results reveal that the flatband is readily destroyed and evolves into a complex dispersive band as a result of the lattice perturbations.

\section{Conclusion}

In summary, we have demonstrated a feasible scheme to realize higher-order EPs (EP3) in flatband rhombic lattices. By introducing on-site gain/loss and non-Hermitian couplings, we have established a diagonal $\mathcal{PT}$-symmetric and an oﬀ-diagonal $\mathcal{PT}$-symmetric rhombic lattices, respectively. Furthermore, we have made careful consideration of the distributions of the higher-order EPs and the dynamical behaviors around them. It has shown that the systems possess both the non-Hermitian CS and SLS, thus leading to the presence of zero-energy flatbands and non-zero EP3. We have also observed varied propagation dynamics in these systems, such as sustained flatband CLSs, exponential and quartic power increase as well as dynamical localization. These results may be useful for realizing the high-sensitivity in a passive wireless sensing system~\cite{zeng2019enhanced,zhang2021demonstration,PhysRevLett.125.240506} and provide insightful information about the underlying properties of the higher-order EPs in flatband systems. In future works, it will also be interesting to investigate a wide range of phenomena such as the topological transitions and nonlinear optical processes in these systems~\cite{PhysRevResearch.4.013195,PhysRevA.103.L040202,PhysRevLett.128.213901,komis2022robustness,PhysRevX.6.021007}. 
\\
\\
\\
\textbf{Acknowledgments:}
This work was supported by the National Key R\&D Program of China (Grant No. 2017YFA0303800), the NSFC (Grant Nos. 11274096, 12134006, 11922408, and 12074105), the 111 Project (Grant No. B07013) in China.

 \hspace*{\fill}
%\appendix
%\begin{appendices}

\appendix

%\section{Appendixes}
\setcounter{equation}{0}
\renewcommand\theequation{B.\arabic{equation}}

\section{The energy amplification in the $\mathcal{PT}$ symmetry broken phase}
\label{APA}
\begin{figure}[h]
	\centering
	\includegraphics[width=0.8\columnwidth]{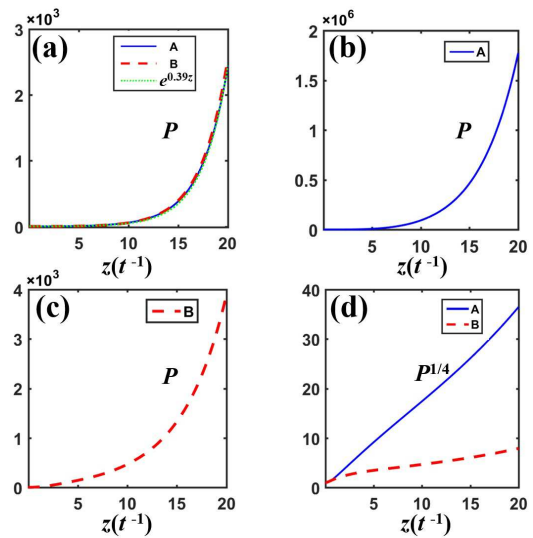}
	\caption{Energy evolution in the symmetry broken phase of the  (a) diagonal and (b-d) off-diagonal  $\mathcal{PT}$-symmetric lattices. (a) The power increase for the excitation of $\bf{A}$ and $\bf{B}$ sublattices are identical and follow an exponential amplification $\left [ P(z) = e^{0.39z} \right ] $. (b-d) The power increases for the excitation of (b) $\bf{A}$ and (c) $\bf{B}$ sublattices follow quartic amplifications $\left [ P(z) \sim z^{4} \right ] $, and the corresponding intensities $P^{1/4} $ increase linearly. Note that the two lines in (d) have different slops.
		\label{figure9}}
\end{figure}

%\noindent
A localized single-site excitation in these $\mathcal{PT}$-symmetric lattices can display exponential and quartic power increases in the $\mathcal{PT}$ symmetry broken phase. Specifically speaking, for the diagonal $\mathcal{PT}$-symmetric lattices, the power increases for the excitation of $\bf{A}$ and $\bf{B}$ sites are identical and follow an exponential amplification $\left [ P(z) = e^{0.39z} \right ] $ (green line in Fig.~\ref{figure9}(a)). For the off-diagonal $\mathcal{PT}$-symmetric lattices, the power increase follows quartic amplifications $\left [ P(z) \sim z^{4} \right ] $. However, the energy amplification is greater for the excitation of $\bf{A}$ site (Fig.~\ref{figure9}(b)). As can be seen in Fig.~\ref{figure9}(d), the corresponding intensities $P^{1/4} $ increase linearly but the two lines have different slopes.

\section{ Eigenvalue evolutions of the $\mathcal{PT}$-symmetric lattices}
%\subsection{Eigenvalue evolutions of the $\mathcal{PT}$-symmetric lattices}
\label{APB}

Here, we provide an analytical study of the eigenvalue evolutions of our $\mathcal{PT}$-symmetric structure in the vicinity of the EP3. For simplicity, we only discuss the case of perturbing the gain sublattice $\bf{A}$. As a result, the perturbation terms only appear along the diagonal element of $H_{k}$. For the diagonal $\mathcal{PT}$-symmetric lattices, the Hamiltonian $H_{k}$  can be written as
\begin{eqnarray}\label{Hamiltonian 1111}
{H_{k}}=\begin{pmatrix}
i\gamma+\epsilon & t_L+t_R{e^{ik} }   & 0\\
t_L+t_R{e^{-ik} }& 0 & t_L+t_R{e^{-ik} }\\
0& t_L+t_R{e^{ik} } &-i\gamma
\end{pmatrix}.
\end{eqnarray}

The corresponding determinant of $H_{k}$ equated to zero is
\begin{eqnarray}
\begin{vmatrix}
-E_{n}+i\gamma+\epsilon  & t_{L}+t_{R} e^{ik}   & 0\\
t_{L}+t_{R} e^{-ik} & -E_{n} & t_{L}+t_{R} e^{-ik}\\
0& t_{L}+t_{R} e^{ik} &-E_{n}-i\gamma
\end{vmatrix}=0.
\end{eqnarray}
The energy bands can be analytically obtained through solving the cubic characteristic equation, which is simplified to
\begin{eqnarray}
0=E_{n} ^{3} -E_{n} ^{2}\epsilon -E_{n}\left ( i\gamma \epsilon -\gamma ^{2} +2A \right )+\epsilon A,
\end{eqnarray}
where $\gamma _{c} -2A=0$, because $A=t_{L}^{2}  +t_{R} ^{2} +2t_{L}t_{R}\cos k=0.16t^{2}$ at the EP3 ($k=\pi, \gamma _{c} /t=2\sqrt{2} g$). By means of Newton–Puiseux series $E_{n} \sim c_{1} \epsilon ^{1/3} +c_{2} \epsilon ^{2/3}$ ($c_1$ and $c_2$ are complex constants), we can expand the above equation by perturbation 

\begin{small}
\begin{eqnarray}
&&0=(c_1^3+A)\epsilon ^{3/3} +(3c_1^2c_2-i\gamma c_1)\epsilon ^{4/3} +(3c_1c_2^2-c_1^2-c_2i\gamma)\notag\\
&&\epsilon ^{5/3}+(c_2^3-2c_1c_2)\epsilon ^{6/3} -c_2^2\epsilon ^{7/3}.
\end{eqnarray}
\end{small}

Forcing the coefficients of the first two terms to be zero, we obtain three sets of values for the coefficients $c_{1}$ and $c_{2}$, corresponding to the three eigenvalues
\begin{eqnarray}
&&E_{-1} \sim \sqrt[3]{\frac{4}{25} } e^{i\pi /3} \epsilon ^{1/3} +\sqrt[3]{\frac{4}{25} } \frac{i\sqrt{2} }{3} e^{-i\pi /3} \epsilon ^{2/3}\notag ,
\\
&&E_{0} \sim \sqrt[3]{\frac{4}{25} } e^{-i\pi /3} \epsilon ^{1/3} +\sqrt[3]{\frac{4}{25} } \frac{i\sqrt{2} }{3} e^{i\pi /3} \epsilon ^{2/3}\notag ,
\\
&&E_{1} \sim -\sqrt[3]{\frac{4}{25} } e^{i\pi /3} \epsilon ^{1/3} -\sqrt[3]{\frac{4}{25} } \frac{i\sqrt{2} }{3} e^{-i\pi /3} \epsilon ^{2/3} .
\end{eqnarray}

Similarly, we can get the results of the off-diagonal $\mathcal{PT}$-symmetric system. The Hamiltonian $H_{k}$ can be written as
\begin{eqnarray}
{ H_{k}}=\begin{pmatrix}
\epsilon & T_L+T_R{e^{ik} }   & 0\\
T_L+T_R{e^{-ik} }& 0 & T^*_L+T^*_R{e^{-ik} }\\
0& T^*_L+T^*_R{e^{ik} } &0
\end{pmatrix}.
\end{eqnarray}
The determinant of $H_{k}$ equated to zero is
\begin{eqnarray}
\begin{vmatrix}
-E_{n}+\epsilon  & T_{L}+T_{R} e^{ik}   & 0\\
T_{L}+T_{R} e^{-ik} & -E_{n} & T_{L}+T_{R} e^{-ik}\\
0& T_{L}+T_{R} e^{ik} &-E_{n}
\end{vmatrix}=0.
\end{eqnarray}
Here, $A=T_L^2+T_R^2+2T_LT_R\cos k=8i$ and $A^{*} =T_L^{*2}+T_R^{*2}+2T_L^*T_R^*\cos k=-8i$ at the EP3 ($k=0, \gamma _{c} /t=1$). So, the determinant can be simplified to
\begin{eqnarray}
0=E_n^3-E_n^2\epsilon+\epsilon A^* .
\end{eqnarray}
Using the Newtonian series expansion for $E_{n} \sim c_{1} \epsilon ^{1/3} +c_{2} \epsilon ^{2/3}$, one can get
\begin{eqnarray}
0=&&(c_1^3-8i)\epsilon ^{3/3} +3c_1^2c_2\epsilon ^{4/3} +(3c_1c_2^2-
c_1^2-c_2i\gamma)\epsilon ^{5/3}\notag\\
&&+(c_2^3-2c_1c_2)\epsilon ^{6/3} -c_2^2\epsilon ^{7/3}.
\end{eqnarray}
The bifurcations in the eigenvalues have the form
\begin{eqnarray}
&&E_{-1} \sim 2e^{i5\pi /6} \epsilon ^{1/3} \notag ,
\\
&&E_{0} \sim 2e^{i\pi /6} \epsilon ^{1/3} \notag ,
\\
&&E_{1} \sim -2i\epsilon ^{1/3} .
\end{eqnarray}
%\subsection{The energy amplification in the $\mathcal{PT}$ symmetry broken phase}

%\\ \hspace*{\fill} \\
\section{The spreading velocity $ \nu $ for the $\bf{B}$ site excitation of the off-diagonal $\mathcal{PT}$-symmetric lattices}
\label{APC}

\begin{figure}[hb]
	\centering
	\includegraphics[width=0.9\columnwidth]{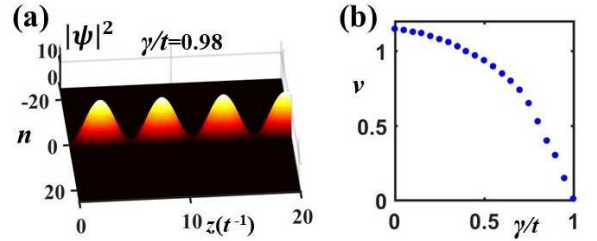}
	\caption{(a) Intensity profile $\left | \psi  \right |^2 $ of a localized $\bf{B}$ sublattice excitation for $\gamma/t=0.98$. When $\gamma/t$ approaches the EP3, the diffraction is prevented. (b) Behavior of the spreading velocity $ \nu $ as a function of $\gamma/t$, computed for the largest propagation length $z$ = 20 $t^{-1}$.  $\nu$ is expressed in units of $t$.
		\label{figure10}}	
\end{figure}

For the off-diagonal $\mathcal{PT}$-symmetric lattices, the $\bf{B}$ site excitation displays a transition from ballistic wave-packet spreading to dynamical localization. As can be seen in Fig.~\ref{figure10}(a)), the diffraction is prevented when the non-Hermitian parameter $\gamma/t$ approaches the value at the EP3. Meanwhile, the spreading speed $ \nu $ decreases faster for large $\gamma/t$ as shown in Fig.~\ref{figure10}(b))

%\end{appendix}

%\begin{backmatter}
%\\ \hspace*{\fill} \\

\textbf{Disclosures:}
The authors declare no conflicts of interest.

\textbf{Data Availability Statement:}
Data underlying the results presented in this paper are not publicly available at this time but may be obtained from the authors upon reasonable request.

%\bmsection{Supplemental document}
%See Supplement 1 for supporting content. 

%\end{backmatter}
%%%%%%%%%% If using BibTeX:
%\\ \hspace*{\fill} \\
\bibliography{apssamp}

%%%%%%%%%% If preparing manually:
% \newcommand{\enquote}[1]{``#1''}

% \bibitem{Zhang:14}
% Y.~Zhang, S.~Qiao, L.~Sun, Q.~W. Shi, W.~Huang, L.~Li, and Z.~Yang,
%   \enquote{Photoinduced active terahertz metamaterials with nanostructured
%   vanadium dioxide film deposited by sol-gel method,}
%   {\protect\JournalTitle{Optics Express}} \textbf{22}, 11070--11078 (2014).

% {Optical Society}, \enquote{{OSA Publishing},}
%   \url{http://www.osapublishing.org}.

% \bibitem{FORSTER2007}
% P.~Forster, V.~Ramaswamy, P.~Artaxo, T.~Bernsten, R.~Betts, D.~Fahey,
%   J.~Haywood, J.~Lean, D.~Lowe, G.~Myhre, J.~Nganga, R.~Prinn, G.~Raga,
%   M.~Schulz, and R.~V. Dorland, \enquote{Changes in atmospheric consituents and
%   in radiative forcing,} in \enquote{Climate Change 2007: The Physical Science
%   Basis. Contribution of Working Group 1 to the Fourth assesment report of
%   Intergovernmental Panel on Climate Change,}  S.~Solomon, D.~Qin, M.~Manning,
%   Z.~Chen, M.~Marquis, K.~B. Averyt, M.~Tignor, and H.~L. Miler, eds.
%   (Cambridge University Press, 2007).

% \end{thebibliography}

\end{document}